\documentclass[10pt,letterpaper]{article}
\usepackage[top=0.85in,left=2.75in,footskip=0.75in]{geometry}

\usepackage{amsmath,amssymb}

\usepackage{changepage}

\usepackage{textcomp,marvosym}

\usepackage{cite}

\usepackage{nameref,hyperref}

\usepackage[right]{lineno}

\usepackage[nopatch=eqnum]{microtype}
\DisableLigatures[f]{encoding = *, family = * }

\usepackage[table]{xcolor}

\usepackage{array}

\newcolumntype{+}{!{\vrule width 2pt}}

\usepackage[T1]{fontenc}
\usepackage[utf8]{inputenc}
\usepackage{bm}
\usepackage[final]{graphicx}
\usepackage{siunitx}
\DeclareSIUnit[number-unit-product = {}]{\bp}{bp}
\DeclareSIUnit{\Mb}{\mega\bp}

\newlength\savedwidth

\raggedright
\usepackage[aboveskip=1pt,labelfont=bf,labelsep=period,justification=raggedright,singlelinecheck=off]{caption}

\makeatletter
\renewcommand{\@biblabel}[1]{\quad#1.}
\makeatother

\usepackage{lastpage,fancyhdr,graphicx}
\usepackage{epstopdf}
\fancyheadoffset[L]{2.25in}
\fancyfootoffset[L]{2.25in}
\newcommand{\vc}[1]{\bm{#1}}
\newcommand{\ie}{\textit{i.e.}}
\newcommand{\nv}{\hat{\bm{n}}}
\newcommand{\missingfigure}[1]{#1}
\newcommand{\activator}{\mathrm{a}}
\newcommand{\complex}{\mathrm{c}}
\newcommand{\substrate}{\mathrm{s}}
\newcommand{\inhibitor}{\mathrm{in}}

\begin{document}
\vspace*{0.2in}

% Title must be 250 characters or less.
\begin{flushleft}
{\Large
\textbf{Diversity in emergent cell locomotion from the coupling cytosolic and cortical Marangoni flows with reaction–diffusion dynamics} % Please use "sentence case" for title and headings (capitalize only the first word in a title (or heading), the first word in a subtitle (or subheading), and any proper nouns).
}
\newline
% Insert author names, affiliations and corresponding author email (do not include titles, positions, or degrees).
\\
Bla\v{z} Iv\v{s}i\'c\textsuperscript{1,2},
Igor Weber\textsuperscript{3},
Piotr Nowakowski\textsuperscript{1*},
Ana-Sun\v{c}ana Smith\textsuperscript{1,4,5*}
\\
\bigskip
\textbf{1} Division of Physical Chemistry, Ru{\dj}er Bo\v{s}kovi\'c Institute, Zagreb, Croatia %Group for Computational Bioscienes
\\
\textbf{2} Centre for Advanced Laser Techniques, Institute of Physics, Zagreb, Croatia
\\
\textbf{3} Division of Molecular Biology, Ru{\dj}er Bo\v{s}kovi\'c Institute, Zagreb, Croatia %Laboratory of Cell Dynamics
\\
\textbf{4} Faculty of Sciences, Friedrich-Alexander-Universit\"at, Erlangen, Bavaria, Germany%PULS Group, Department of Physics, Center for Computational Advanced Materials and Processes
\\
\textbf{5} Competence Center Engineering of Advanced Materials, Friedrich-Alexander-Universit\"at, Erlangen, Bavaria, Germany %Center for Computational Advanced Materials and Processes
\\
%\textbf{3} Affiliation Dept/Program/Center, Institution Name, City, State, Country
%\\
\bigskip

% Insert additional author notes using the symbols described below. Insert symbol callouts after author names as necessary.
% 
% Remove or comment out the author notes below if they aren't used.
%

% Use the asterisk to denote corresponding authorship and provide email address in note below.
* Piotr.Nowakowski@irb.hr, asmith@irb.hr, ana-suncana.smith@fau.de

\end{flushleft}
% Please keep the abstract below 300 words

% For PLOS Medicine research article authors, please structure your abstract
% with "Background", "Method and Findings" and "Conclusion" sections per
% journal requirements.

% For PLOS Neglected Tropical Diseases research article authors, please
% structure your abstract with "Background", "Methodology", "Findings", and
% "Conclusion" sections per journal requirements.
%
\section*{Abstract}
Cell migration is a fundamental process underlying the survival and function of both unicellular and multicellular organisms. Crawling motility in eukaryotic cells arises from cyclic protrusion and retraction driven by the cytoskeleton, whose organization is regulated by reaction–diffusion (RD) dynamics of Rho GTPases between the cytosol and the cortex. These dynamics generate spatial membrane patterning and establish front–rear polarity through the coupling of biochemical signalling and mechanical feedback. We develop a cross-scale mean-field framework that integrates RD signalling with cytosolic and cortical hydrodynamics to capture emergent cellular locomotion. Our model reproduces diverse experimentally observed shape and motility phenotypes with small parameter changes, indicating that these behaviours correspond to self-organized limit cycles. Phase-space analysis reveals that coupling to both cytosolic flow and spatially varying surface tension is essential to recover the full spectrum of motility modes, providing a theoretical foundation for understanding amoeboid migration.

% Please keep the Author Summary between 150 and 200 words. Use first person.
% PLOS ONE, PLOS Biology, PLOS Global Public Health, PLOS Mental Health, and PLOS Water authors please skip this step. Author Summary is not valid for submissions to these journals.

% For PLOS Medicine authors, please structure your author summary with answers to the following questions:
% Why was this study done?
% What did the researchers do and find?
% What do these findings mean?
%
\section*{Author summary}
How do free-living amoeboid cells decide where to go when no clear external cues guide them? We developed a computer simulation that mimics how simple cells move freely across a flat surface. The model connects simple chemical reactions within the cell cortex with the flow on the cell surface and of its inner fluid. In this way, the shape of cell changes and locomotion is induced. Our simulations show that this internal coordination alone can produce motion patterns seen in real cells---steady gliding, shifting directions, circling, or turning intermittently. By tuning key coupling parameters, we found how small internal changes can trigger very different behaviours. The study hence reveals how, based on simple physical principles, cells can self-organise and explore their surroundings.

\newpage
%\linenumbers

% Use "Eq" instead of "Equation" for equation citations.
\section*{\label{sec:Intro}Introduction}

Cell migration is a fundamental biological process essential for the survival of both unicellular and multicellular organisms, spanning from flagellated bacteria navigating liquid environments to motile fibroblasts orchestrating wound closure in animals. Crawling motility in eukaryotic cells relies on cyclic protrusion and retraction driven by the cytoskeleton, a composite network of actin filaments, microtubules, and intermediate filaments~\cite{Sackmann2010, MerinoCasallo2022}. Coordination among these structural components is mediated by a complex signalling network, prominently the Rho family of GTPases, which act as molecular switches regulating cytoskeletal dynamics~\cite{Mosaddeghzadeh2021}. In their active, GTP-bound state these proteins transmit signals to downstream effectors that locally modulate filament assembly and contractility~\cite{Jaffe2005}. Even in the absence of external stimuli, Rho GTPases can self-organize through reaction–diffusion (RD) dynamics operating between the cytosol and the cell cortex~\cite{Halatek2018,Sostar2024,Bement2024}. Such dynamics give rise to spatial membrane patterning with localized zones of high activator concentration that promote actin polymerization and define cell polarity. The interplay between cortical and cytosolic flows, together with RD feedback, establishes a tightly coupled system in which biochemical signalling and mechanical deformation co-regulate cell shape, polarity, and motility in a continuously self-organizing manner.

Numerous approaches have been proposed to model such a complex biological task~\cite{Holmes2012,Ziebert2016,Buttenschon2020}. A variety of approximations and computational techniques can be employed to model each of the aforementioned processes with varying levels of biological complexity and precision. Main difference among these approaches lies in the way forces are incorporated with the RD system and in the way interface and the position of the cell is tracked. Before outlining the methods available for tracking shape and translating the cell body, we briefly introduce the approach to model signalling dynamics.

Since Turing's pioneering work~\cite{Turing1952}, which demonstrated that two interacting and diffusing substances can form stable spatial patterns in their concentration distribution, RD systems have become an indispensable approach for modelling numerous biological processes. Moreover, RD systems extend Turing's original idea by revealing not only stationary but also time-dependent dynamics, which can explain cyclic biological processes~\cite{Beta2017}. It has been shown repeatedly that RD systems, based on known interactions of the Rho family of GTPases and related proteins, can explain stable polarization~\cite{Jilkine2007, Otsuji2007} as well as oscillatory dynamics~\cite{Bement2015, Chiou2018, Sostar2024}.

The aforementioned dynamics directly influence how the cytoskeleton is organized. To model the motility process, one thus needs to couple RD system to a component which tracks and translates the cell. The simplest way of approaching this task is to track only the cell contour on a two-dimensional domain. Such model was proposed by Nielson~\textit{et al.}~\cite{Neilson2010}. The authors combine Parametrized Finite Element Method (PFEM), used to track and evolve the interface, with a simple RD system which governs the normal velocity of the cell membrane. The main drawback of such an approach is the fact that tangential evolution of the nodes, which form the contour of cell, is controlled by Moving Mesh Partial Differential Equations (MMPDEs), which do not necessarily align with realistic biological mechanics.

Apart from tangential evolution, another shortcoming of this approach is the lack of cytoplasmic dynamics inside the cell. To address this, two upgraded models similar to PFEM were proposed~\cite{MacDonald2016, Mackenzie2019}. Both approaches employ explicit bulk and surface tracking by evolving the positions of nodes that define the interior, exterior, and cortex of the cell. The main advantage of these so-called Arbitrary Lagrangian-Eulerian (ALE) approaches is their reduced computational cost compared to other high-resolution models, although this comes at the expense of a mathematically complex formulation of the MMPDEs governing node translation. However, ALE-based models struggle to fully capture membrane tension and intracellular forces that influence the bulk nodes. Instead, these models resemble mechanical spring-force approximations.

Another type of models that usually incorporate the RD dynamics with the changes in cell shape better than the previously mentioned ones are Cellular Potts Models (CPM). In these models, Monte Carlo algorithm is used to evolve the cell interface. Maree~\textit{et al.}~\cite{Maree2012} implemented CPM with a highly complex signalling network, incorporating multiple protein species and their interactions. This highlights the primary strength of the CPM approach: it can be easily coupled with intricate biochemical networks to model a wide range of biophysical processes involved in cell motility, while maintaining a relatively low computational cost. However, a key limitation of CPM is that the shape evolution dynamics are inherently stochastic and do not necessarily reflect the physical nature of the forces driving cell motility.

To address the stochastic nature of shape changes in CPM, one can employ a Phase Field (PF) model. This class of models naturally integrates free energy formulations into their evolution equations, improving the physical description of the system, and has been widely used for simulating cell motility~\cite{Aranson2016, Shao2012, Alonso2018, Moreno2020, Moreno2022a}. PF models fall into the category of diffuse interface models, meaning that the cell is represented by a smooth phase-field function that transitions gradually across the simulation domain. On the downside, the diffuse interface imposes a lower limit on grid spacing and resolution to ensure accurate representation of the phase-field transition from the cell interior to its exterior.

The drawback of all of the aforementioned models lies in the fact that they do not account for the flow inside the cytoplasm which can affect the RD system in moving cells. This can be mitigated by the use of an approach similar to PF called the Level-Set (LS) method. Originally developed for simulating phase flow problems~\cite{Osher1988}, this technique has been effectively applied to modelling of cell migration. Within LS formalism, the cell boundary is naturally embedded within a continuous field. Notably, this formulation assumes that the cortex is a one-dimensional curve of zero thickness. Several studies have successfully employed LS methods for cell motility simulations~\cite{Kuusela2009, Shi2013, Schindler2024}, each incorporating different techniques to model signalling dynamics. The main advantage of the LS approach, particularly compared to the previously mentioned methods, is its ability to simulate highly irregular cell shapes and their dynamic deformations. However, a key drawback is the gradual degradation of the LS function over time.

The explicit interface tracking of the LS method can itself be challenging, especially when trying to incorporate RD systems that have defined surface protein concentration and their diffusion. An alternative approach~\cite{Moure2016}, which combines elements of both LS and PF methods, is to define the LS as smooth function across the interface (such as a hyperbolic tangent), rather than a step or a signed distance change.
Although the interface position is still determined by a predefined contour level, it is now diffuse rather than sharp.

In this paper, we propose a minimal model of cellular locomotion that combines different aspects of the aforementioned approaches. To account for cytosolic flow, cell shape, and RD dynamics~\cite{Bruckner2024}, we link the LS method with the Navier-Stokes equation to track cell position and account for cytosolic and cortex flows, and couple these components to the canonical RD system. This allows for simulation of non-stimulated cell motility as well as chemotactic locomotion (with minimal updates to the model).  

\section*{\label{sec:Model}Model}

\begin{figure}[h!]
\missingfigure{\includegraphics[width=0.7\textwidth]{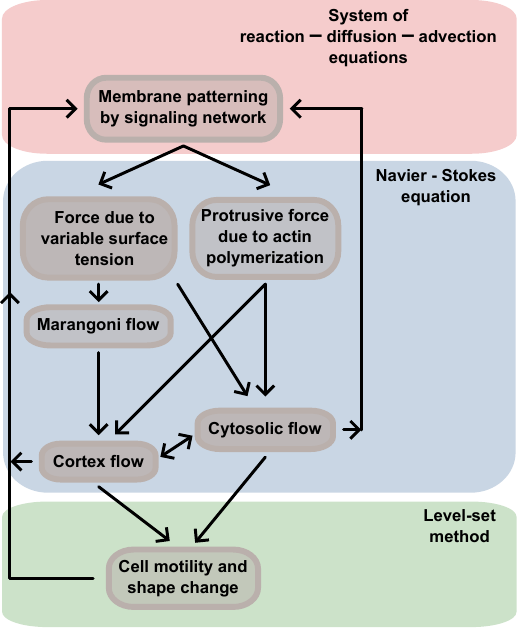}}%
\caption{\label{fig:modelscheme} \textbf{Diagram of the components of the model.} To fully capture complex biological process of cell motility one needs to couple membrane patterning by reaction--diffusion--advection network to forces which induce cytosolic and membrane flows as well as facilitate shape change and motility in general. The coupling loop is closed by allowing the generated flows and geometry of the cell to influence reaction--diffusion--advection system.}
\end{figure}

We propose a model of cell locomotion that combines LS formalism with fluid flow and reaction--diffusion--advection dynamics. The components of the model and the relations between them are shown schematically in Fig~\ref{fig:modelscheme}.

\subsection*{\label{sec:LS}Level-set formalism}

Motivated by two-phase-flow applications, we adopt a LS approach, specifically a diffuse-interface variant~\cite{Osher2003}. This method allows for separation of different regions with a scalar field $\phi$, referred to as LS function. The evolution of this function is governed by the conservative continuity equation~\cite{Osher1988}
\begin{equation}
\frac{\partial\phi}{\partial t}+\vc{\nabla}\cdot(\vc{v}\phi)=0,
\label{eqn:LS1}
\end{equation}
which transports~$\phi$ under the prescribed velocity field~$\vc{v}$, and we use $t$ to denote time.

Given the freedom to specify the shape of the LS function, we adopt a
diffuse-interface variant, common in phase-field (PF) models~\cite{Anderson1998}. We model the shape by demanding $\phi=1$ inside and $\phi=0$ outside the cell, with a continuous smooth function connecting these two regions in the shape of hyperbolic tangent~\cite{Anderson1998,Moure2016}
\begin{equation}
\phi\left(r,\theta\right)=\frac{1}{2}\left[1-\tanh\left(\frac{r-r_0\left(\theta\right)}{\epsilon/3}\right)\right],
\label{eqn:LS2}
\end{equation}
where, for the sake of clarity, we have used polar coordinates $\left(r, \theta\right)$, and assumed that the location of the interface can be described by an angle-dependent distance as $r_0(\theta)$. The parameter $\epsilon$ controls the width of the crossover.

Employing a hyperbolic-tangent profile \eqref{eqn:LS2} creates an interface region of thickness $\epsilon$ which allows us to model membrane processes within our two-dimensional framework. To this end, we define~\cite{Peskin2002}
\begin{equation}
\delta_\epsilon\left(\vc{r}\right)=\left|\vc{\nabla} \phi \right|
\label{eqn:LS3}
\end{equation}
as a numerical approximation of the Dirac delta function supported on the interface line $\phi=1/2$, where the vector $\vc{r}$ denotes a point on the plane. Since the integral of $\delta_\epsilon$ in the direction perpendicular to the interface is $1$, this function can be used to localise biochemical processes in the cell cortex. Furthermore, in the diffuse-interface framework geometric quantities, such as normal vector or interface curvature, can be derived from derivatives of~$\phi$~\cite{Osher2003} (see Sec~1 of \nameref{S1_Appendix} for details).

A well-known drawback of this approach is a slow degradation of the level set function profile by advection (Eq~\eqref{eqn:LS1})~\cite{Russo2000}. We restore the desired profile by repeatedly performing reinitialisation, \ie,~by evolving~$\phi$ in a pseudo-time~$\tau$ between physical time steps, using the conservative scheme proposed by Parameswaran and Mandal~\cite{Parameswaran2023}.

\subsection*{\label{sec:FD}Fluid dynamics}

For the evolution of the velocity field $\vc{v}$ we use the Navier--Stokes equation~\cite{Landau1987}
\begin{equation}
\frac{\partial\vc{v}}{\partial t}=-\frac{1}{\rho}\vc{\nabla}p+\nu\vc{\nabla}^2\vc{v}+\vc{f},
\label{eqn:FD1}    
\end{equation}
where $\rho$ is the density of the fluid, $\nu$ is the kinematic viscosity, $p$ is the pressure, and $\vc{f}$ denotes external force acting on the fluid. In the equation above, we neglect the non-linear term as we are interested in low-Raynolds-number flows, and we assume that the liquid is incompressible ($\vc{\nabla}\cdot\vc{v}=0$) which determines the pressure $p$.

We assume that the force acting on the fluid can be decomposed as~\cite{Brackbill1992,Whitfield2016,Mogilner2003,Barnhart2011,Fogelson2015}
\begin{equation}
\vc{f}=\left[\sigma\kappa\nv+\vc{\nabla}\sigma-\nv\left(\nv\cdot\vc{\nabla}\sigma\right)+k_\mathrm{C} C_{\activator}\nv
-\alpha\left(A-A_0\right)\nv\right]\delta_\epsilon\left(\vc{r}\right)-\beta\vc{v}.
\label{eqn:FD2}
\end{equation}
where $\sigma$ denotes surface tension, $\nv=-\vc{\nabla}\phi/\left|\vc{\nabla}\phi\right|$ is a unit vector normal to the interface, $\kappa=\vc{\nabla}\cdot \nv$ is the local curvature of the interface, $C_\activator$ is the concentration of activator species (see below), and $A$ denotes the area of the cell. In the above equation $k_\mathrm{C}$, $\alpha$, $A_0$, and $\beta$ denote parameters of the model which we discuss below, and the factor $\delta_\epsilon\left(\vc{r}\right)$ ensures that forces described by all terms but last only act on the interface.

In Eq~\eqref{eqn:FD2} the first term ($\sigma\kappa\nv$) describes the normal force resulting from the surface tension $\sigma$. We assume a linear dependence
\begin{equation}
\sigma=\sigma_0+k_\sigma C_{\activator},
\label{eqn:FD3}
\end{equation}
such that the surface tension is modified from its base value $\sigma_0$ by a term proportional to the concentration of the activator $C_\activator$ with the coefficient $k_\sigma$. The space-dependent surface tension induces the Marangoni surface flow \cite{Chen2015, Schmitt2016, Whitfield2016} (from regions of lower to higher surface tension) described by the second and third terms ($\vc{\nabla}\sigma-\nv\left(\nv\cdot\vc{\nabla}\sigma\right)$) in Eq~\eqref{eqn:FD2}. 

The idea of spatially dependent cortical tension generated by the network of actin filaments has been well established in the literature \cite{Merkel2000, Clark2014, Chugh2017, Li2024, GarciaArcos2024}. However, to the best of our knowledge, its effect on cortical flow has not been studied numerically in the context of cell motility.

The fourth term in Eq~\eqref{eqn:FD2} ($k_\mathrm{C} C_\activator\nv$) represents the protrusive force generated by actin polymerization at the leading edge~\cite{Mogilner1996, Mogilner2003}. We assume that this force is proportional to the concentration of activator $C_\activator$ with coefficient $k_\mathrm{C}$. The fifth term ($-\alpha\left(A-A_0\right)\nv$) enforces the near-constancy of the projected cell area $A\approx A_0$, as observed experimentally~\cite{Barnhart2011}. Finally, the linear drag term $-\beta\vc{v}$ models the dissipation and ensures that the flow relaxes when external forcing stops~\cite{Fogelson2015}.

Before proceeding, we briefly justify the specific structure of the force density in Eq~\eqref{eqn:FD2}. First, the Marangoni flow can strongly bias the otherwise diffusive transport of membrane-bound proteins ~\cite{Scriven1960,Sens2015}. Second, letting both surface tension and protrusive force depend on local activator concentration directly couples cortical mechanics to the biochemical polarity cue: high activator activity marks the protruding edge of the cell~\cite{Machacek2009,Houk2012}. Because the leading and trailing edges exhibit different membrane tensions~\cite{Houk2012,Whitfield2016}, we represent relative changes in $\sigma$ with the activator-dependent law~\eqref{eqn:FD3} rather than adding further geometric factors. Finally, the actin-generated protrusive force is approximated to act normally on the cortex, consistent with the mean orientation of the filament polymerization against the membrane~\cite{Mogilner2003}. Eqs~\eqref{eqn:FD1}--\eqref{eqn:FD3} thus provide a minimal, yet mechanistically grounded, route by which the distribution of a polarity protein modulates the cell shape as well as intracellular and surface flows.

\subsection*{\label{sec:RDA}Reaction--diffusion--advection system}

The objective of the reaction--diffusion--advection (RDA) model is to capture the dynamics of an activator whose concentration is coupled to the force profiles in the cell cortex and to the substrate from which the activator is produced~\cite{Turing1952}. These coupled species constitute the core of a minimal Turing-type activator--substrate model capable of producing stationary polarity patterns~\cite{Bement2015, Goryachev2019}. To extend the available solution space to include time-dependent (Hopf-type) and wave-pinning solutions, we need to introduce at least one additional protein species. Moreover, at least four interacting protein species are required to reproduce all of the above solution types~\cite{Sostar2022}. 

In the current work, we used \textit{Dictyostelium discoideum} as an example system for which RD equations have been shown to reproduce experimentally observable activator dynamics in stationary cells~\cite{Sostar2024}. We opted for a reduced version of the original model consisting of cytosolic substrates, membrane-bound activator, and a complex between the activator and its inhibitor. Such interaction network can be easily related to the Rho GTPase switching cycle~\cite{Das2012,Graessl2017,Michaux2018,Bement2024}. For example, the role of the activator can be attributed to the active form of a GTPase, while its cytosolic substrate can be related to its inactive form bound to RhoGDI. Biologically, the signalling cycle ends with the deactivation of the GTPase mediated by an inhibitor protein, GAP, and in our model this interaction is modelled by formation and dissociation of the activator--inhibitor complex.

Based on this rationale, we define four protein species. Two are cytoplasmic, \ie, restricted to the bulk region, and two are constrained to the cell cortex. The membrane species that plays a role of the activator that couples to the forces is denoted by the index ``$\activator$''. The bulk species, that acts as the substrate for this activator, is denoted by index ``$\substrate$''. These two species comprise a minimal Turing-type model. To expand the solution space with wave-pinning-type solutions, we need to add a delayed inhibition. This is achieved by introducing a second membrane species that also catalytically promotes the formation of the activator, which we denote by the index ``$\complex$'', representing a complex formed from the activator and another protein. The partner required to form these complexes is an inhibitor species restricted to the bulk, which we denote by the index ``$\inhibitor$''.

\begin{figure}[h!]
\missingfigure{\includegraphics[width=0.9\textwidth]{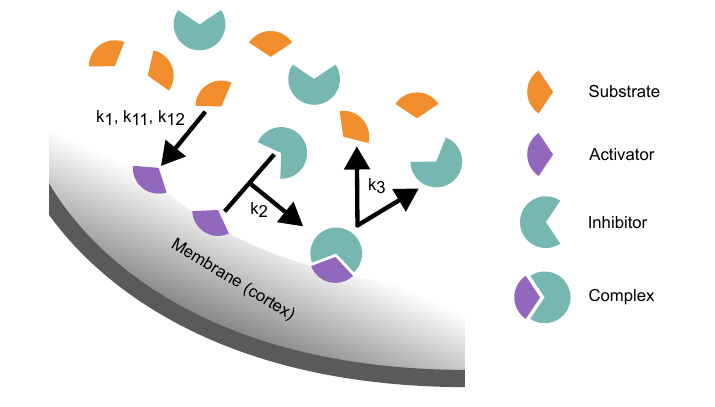}}%
\caption{\label{fig:reactions} \textbf{Schematic representation of the reactions of the model.} Arrows indicate the direction of the reaction with governing constants for that reaction annotated beside them. Cytosolic substrate binds to the membrane becoming the activator. Membrane bound activator can form a complex with the cytosolic inhibitor. The cycle ends when the complex dissociates, releasing substrate and inhibitor back into the cytosol.}
\end{figure}

We consider three reactions of the species: First, a cortex-bound activator ``$\activator$'' is produced autocatalytically from the bulk substrate ``$\substrate$''. Second, the activator ``$\activator$'' together with the bulk inhibitor ``$\inhibitor$'' can form a membrane-bound complex ``$\complex$''. Finally, the complexes can dissociate back into two bulk species. These reactions are presented schematically in Fig~\ref{fig:reactions}. Each species diffuses freely, either two-dimensionally in the bulk cytoplasm or, effectively, one-dimensionally along the membrane cortex, and the reactions are thus restricted to the overlap between the diffuse-interface region and the bulk. For each of the species we introduce a bulk concentration field $C_{i}\left(\vc{r},t\right)$, where $i=\activator, \complex, \substrate, \inhibitor$ denotes the type of species. We use the following mass-conserving reaction terms for each of the species in the RDA equations:
\begin{subequations}\label{eqn:RDA1}
\begin{align}
R_{\substrate} &= k_3C_{\complex}- C_{\substrate}\left(1-\frac{C_\activator}{C_\activator^{\text{MAX}}}\right)\left(k_1 + k_{11}C_\activator + k_{12}C_\complex\right), \\
R_{\inhibitor} &= k_3C_\complex -k_2C_\activator C_{\inhibitor},\\
R_\activator&= -k_2C_\activator C_{\inhibitor}+ C_{\substrate}\left(1-\frac{C_\activator}{C_\activator^{\text{MAX}}}\right)\left(k_1 + k_{11}C_\activator + k_{12}C_\complex\right),\\
R_{\complex} &= k_2C_{\activator}C_{\inhibitor}- k_3C_{\complex},
\end{align}
\end{subequations}
where we have introduced several reaction rates: $k_{1}$ is the basal rate at which the cytosolic substrate ``$\substrate$'' binds and activates on the membrane, $k_{11}$ and $k_{12}$ quantify the co-operative autocatalysis mediated, respectively, by the membrane activator ``$\activator$'' and the transient complex ``$\complex$''. The positive feedback provided by $k_{11}$ supplies the classical activator self-activation mechanism of Turing-type systems~\cite{Turing1952,Goryachev2019}. The additional delayed feedback through $k_{12}$ (activator ``$\activator$'', that is momentarily inactive within the complex ``$\complex$'', still promotes further activator ``$\activator$'' recruitment) has been shown to broaden the parameter range that supports pattern formation in mass-conserved models~\cite{Otsuji2007,Mori2008}. Complex formation and dissociation are governed by $k_{2}$ and $k_{3}$, respectively. Finally, the factor $\left(1-C_\activator/C_\activator^\text{MAX}\right)$ prevents the activator ``$\activator$'' from exceeding a limit of steric saturation on the membrane~\cite{Douglass2005,Mori2008}.

The postulated reaction rates allow us to write the RDA equations for all species:
\begin{equation}
\frac{\partial C_i}{\partial t}+\vc{\nabla}\cdot\left(\vc{v}C_i\right)=
\vc{\nabla}\cdot\left(D_i\vc{\nabla}C_i\right)-
\vc{\nabla}\cdot\left(\frac{D_i}{k_B T}\vc{F}_i C_i\right)+R_i,
\label{eqn:RDA2}
\end{equation}
such that each concentration field $C_i$ is advected with velocity $\vc{v}$, diffuses with constant $D_i$, interacts with external force $\vc{F}_i$, and is subject to reactions given by $R_i$ in Eq~\eqref{eqn:RDA1}. In the above equation, the index $i=\activator, \complex, \substrate, \inhibitor$ denotes the type of species, $k_\mathrm{B}$ is the Boltzmann constant, and $T$ is the temperature.

The role of the introduced forces $\vc{F}_i$ is to restrict surface and bulk species to the desired domains. To this end, we use $\vc{F}_i=-\vc{\nabla}U_i$, and assume
\begin{subequations}\label{eqn:RDA4}
\begin{align}
U_{\substrate}=U_{\inhibitor}&=-k_BT\ln(\phi\left(\vc{r}\right)),\\
U_\activator=U_\complex&=-k_BT\ln\left(\delta_\epsilon\left(\vc{r}\right)\frac{2\epsilon}{3}\right),
\end{align}
\end{subequations}
such that in the equilibrium, without reactions, the probability distribution becomes proportional to $\phi\left(\vc{r}\right)$ or $\delta_\epsilon\left(\vc{r}\right)$ for bulk or surface species, respectively. We have chosen the confining-force method of restricting species over the pointwise Dirichlet constraints because this method strictly preserves mass conservation of each of the species and remains compatible with the diffuse-interface formulation of LS function.

Finally, we note that reactions between cytosolic and membrane species are possible only in the finite overlap zone of interface and bulk confining potentials. The reaction rates in our model can be related to those in models with a one-dimensional interface \cite{Sostar2024} by integrating the local kinetics across the overlap region, a procedure analogous to the surface--volume coupling used in diffuse-interface models of phase flows with soluble surfactants~\cite{Teigen2011}.

\section*{\label{sec:Meth}Methods}

To analyse our model, we have solved the relevant equations numerically using well established methods of computational physics~\cite{Harlow1965,Chorin1968,Moure2016,Oono1988}. We considered a square region with periodic boundary conditions in which we discretised all relevant fields of our model. Then, all spatial derivatives were replaced with finite differences and we advanced system in time using forward-Euler time integration procedure. The details of this procedure are presented in Sec~1 of \nameref{S1_Appendix}.

The calculations were performed using a specially written computer programme in ISO~\texttt{C}99~\cite{C99}. The parameters were chosen to follow the experimental measurements and the numerical works studying \textit{Dictyostelium discoideum} (\textbf{here insert relevant references}). In the simulations we varied coupling constants $k_\mathrm{C}$ and $k_\sigma$, as well as diffusion constants $D_\activator$ and $D_\complex$. The complete list of the values of the parameters used in the simulations is presented in \nameref{S1_Appendix}.

Special attention was paid to the initial configuration of the concentration fields for all species. To this end, we performed preliminary simulations in which the circular shape of the cell was fixed by blocking the evolution of the LS and velocity fields. As we observed, in this case, the RD dynamics can generate three different patterns of proteins: oscillations, rotations, and stable polarisation. To speed up our simulations, we have used the oscillatory and rotational patterns established in the preliminary simulations as initial configurations. The details of the preliminary simulations are discussed in Sec~2 of \nameref{S1_Appendix}.

The analysis of the results was performed using commercially available programmes~\cite{Fiji, Quimp, MATLAB2023b}. By finding the centre of mass for each snapshot, we determined the \textit{velocity} of the cell and plotted its \textit{trajectory}. Cell \textit{elongation} was calculated as $E=\log_2\left(a/b\right)$, where $a$ and $b$ denote the axes of an ellipse fitted to the cell shape~\cite{Dunn1990, Teague1980}. The profile of the activator concentration $C_\activator$ was quantified by the \textit{polarity vector}. In the case of a single activator patch in the cortex, this vector points from the centre of the cell toward the patch. Finally, to study the \textit{flow field}, we transformed the velocity $\vc{v}$ to the centre-of-mass reference frame and, using the LS function, split the flow into surface and cytosolic parts. The details of our analysis are reported in Sec~1 of \nameref{S1_Appendix}.

\section*{\label{sec:Prop}Properties of the solutions}

\subsection*{\label{sec:Flows}Flows induced by protrusive coupling and varying surface tension}

We start analysis of the solutions of our model by studying the effect of the activator on the flow. In Eq~\eqref{eqn:FD2} there are two terms dependent on $C_\activator$: the protrusive force with coupling $k_\mathrm{C}$, and the surface tension forces via concentration-dependent $\sigma$ (see Eq~\eqref{eqn:FD3}) with a coupling constant $k_\sigma$.

\subsubsection*{\label{sec:FlowProt}Protrusive coupling flow}

\begin{figure}[h!]
\missingfigure{\includegraphics[width=0.98\textwidth]{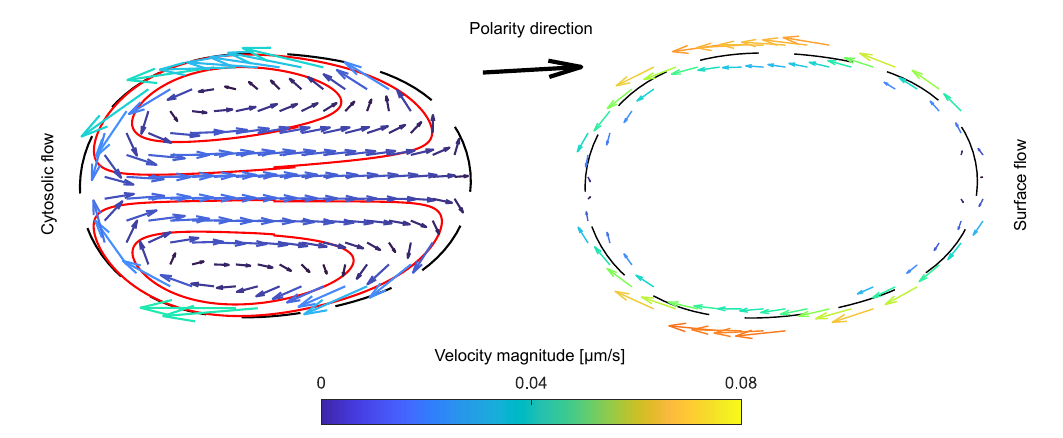}}%
\caption{\label{fig4} \textbf{Flow induced by protrusive coupling.} Cytosolic and cortex flows, in the centre-of-mass reference frame of the cell, during movement under the influence of only protrusive coupling for $k_\mathrm{C}=0.04$ and $k_\sigma=0$. The black arrow marks the polarity direction (and, consequently, the direction of motion as well as the position of the patch of activator). Red lines represent streamlines inside the cytosol. The colour scale for both plots is shown in the middle. Left panel shows cytosolic flow and right panel presents cortex flow induced by protrusive coupling.}
\end{figure}

We first study the effect of protrusion forces that are modelling the effects of actin filaments in real cell. To this end, we have initiated a series of simulations with $k_\mathrm{C}>0$ and $k_\sigma=0$ (\ie, with constant surface tension $\sigma=\sigma_0$). We started from a circular shape of the cell and used an oscillatory initial protein concentration profiles. Under these conditions, when $k_\mathrm{C}$ is large enough, the simulated cell elongates into elliptical shape and moves in a fixed direction defined by the location of the activator patch on the cortex. In Fig~\ref{fig4} we present a plot of a typical flow observed in this case.

As shown in the top panel of Fig~\ref{fig4}, inside the cytosol the flow forms two vortices---the fluid is carried from the front (right side of the plot) to the rear of the cell along the top and bottom part of the cortex, the two flows meet at the back and turn, transporting fluid from the rear to the front along the long axis of cell. Furthermore, inspection of the streamlines reveals that particles trapped within the vortices seldom leave the closed loops along which they circulate. This observation is in a full agreement with both experimental and previously reported numerical studies~\cite{Niwayama2016, Htet2025}.

In the bottom panel of Fig~\ref{fig4} we show the cortex flow. It is strongly affected by the movement of the cell: in the reference frame of the cell, the fluid streams along the cortex, inducing stress as it is dragged along the interface. The resulting cortex flow runs from the front to the rear of the cell along the perimeter with a varying magnitude---low in front and back of the cell and high in the middle regions.

The induced cortex flow carries cortical protein species along the perimeter from the position of the patch of activator. As a result, the diffusive transport of proteins from the front to the rear is enhanced. At the same time, the cytosolic flow inside transports substrate species from rear directly to the front along the long axis of the cell, facilitating the polarization and maintaining the patch. This mechanism facilitates static polarization and subsequent persistent motility~\cite{Illukkumbura2020, Meindlhumer2023, Wigbers2020}.

\subsubsection*{\label{sec:FlowST}Surface tension coupling flow}

Motile cells are typically polarized, exhibiting a front-rear asymmetry in cortex mechanics (often higher cortical tension at the rear for cell-body retraction) while advancing via actin-based protrusions such as filopodia or pseudopodia at the front~\cite{Chugh2018}. To capture this behaviour, our model includes a direct coupling between activator concentration $C_\activator$ and cortex tension $\sigma$ regulated by the parameter $k_\sigma$ (see Equation\,\eqref{eqn:FD3}), which may be both positive or negative, leading to systems with either higher or lower frontal cortex tension, respectively.

The flow arising from variable surface tension, \ie, the Marangoni flow, is directed from regions of lower surface tension toward regions of higher surface tension. Consequently, for positive coupling, the flow is directed toward the maximum activator concentration, while for negative coupling, it is directed in the opposite direction. We refer to these two cases as a \textit{constrictive} flow (for $k_\sigma > 0$) and \textit{dispersive} flow (for $k_\sigma < 0$), respectively.

\begin{figure}[h!]
\missingfigure{\includegraphics[width=0.98\textwidth]{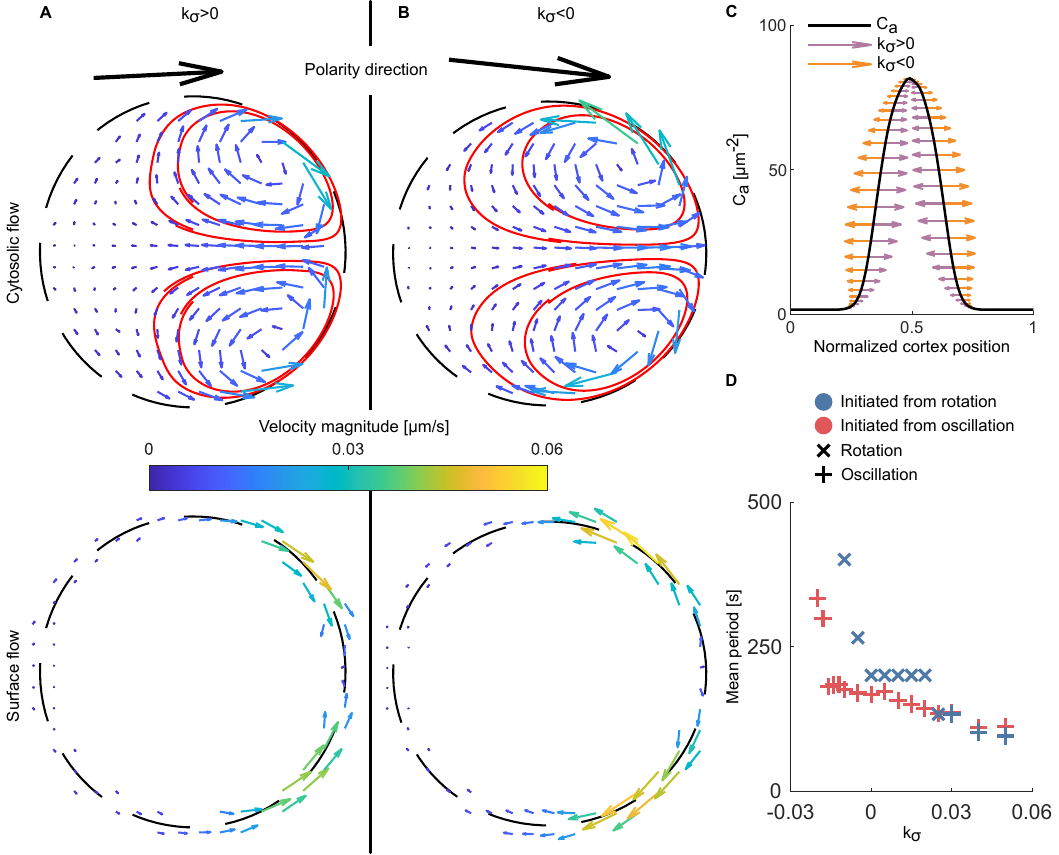}}%
\caption{\label{fig5} \textbf{Constrictive and dispersive flow.} The flow induced in the cell without protrusion force ($k_\mathrm{C}=0$) \textbf{A:} for a positive surface tension coupling $k_\sigma=0.03$, and \textbf{B:} for a negative coupling $k_\sigma=-0.03$. On both panels the patch of activator is located on the right side of cell and the snapshot has been taken in the initial phase of the simulation, where the effect of Marangoni flow is the best visible. In both panels the flow field is presented with arrows and streamlines are denoted with red lines. The magnitude of velocity is also shown with the colour code as presented on the scale bar in the middle of panels. Top and bottom graphs show the cytosolic and the cortex flows, respectively. \textbf{C:} The effect of Marangoni flow on the profile of activator concentration along the cortex. Purple arrows refer to positive surface tension coupling while orange ones represent negative surface tension coupling. \textbf{D:} Mean period of the rotation or oscillation RD dynamics as a function of surface tension coupling. Colour marks the type of initial concentration profile for each simulation. Markers represent the observed limit cycle RD pattern obtained after \SI{1300}{\second} of simulation time. Mean period was averaged over \SIrange{900}{1300}{\second} of simulation time. Higher values of surface tension coupling speed up the RD dynamics and favour oscillation pattern. After the threshold of $0.03$ is passed limit cycle solution of simulations initiated with rotation pattern becomes oscillation.}
\end{figure}

To analyse in detail the effect of this coupling, we initiated two simulations starting from a circular cell with oscillatory initial concentration profiles without protrusion forces ($k_\mathrm{C}=0$) and with positive and negative $k_\sigma$, respectively. The typical flows induced inside the cytosol and along the cell cortex are shown in Fig~\ref{fig5}. As these cells do not necessarily form static polarization patterns, to illustrate the properties of flow profiles in the best possible way, the presented snapshots are from initial stages of simulations, before the flow manages to deform the shape of the cortex and other mechanisms become relevant.

In Fig~\ref{fig5}A, we present the constrictive flow induced in the cytosol and in the cortex for positive coupling $k_\sigma>0$. The Marangoni flow carries fluid along the cortex toward the activator patch at the front of the cell (right side of cells in Fig~\ref{fig5}). The fluid cycles back to the rear along the $x$-axis, forming two vortices. The flow is the strongest in the cortex near the top and bottom boundary of the patch of activator, where the gradient of surface tension is the strongest. In contrast with the case of protrusive flow (see Fig~\ref{fig4}), in the present case, the streamlines are concentrated at the front of the cell, indicating that transport in the rear part is negligible. For the dispersive case of negative coupling $k_\sigma<0$, as shown in Fig~\ref{fig5}B, the flow is naturally reversed: fluid moves toward the front along the $x$-axis and returns to the rear along the cortex. However, the streamlines reveal that the overall characteristics remain the same with two vortices located at the front of the cell. The consequence of localised constrictive or dispersive flow is that the rear of the cell becomes largely segregated from the dynamics of the system as the transported proteins can bypass the rear part of the cell in their cycling.

Despite their similarity, these two cases differ in the direction of the flow which makes the dynamics of the system inherently different. In Fig~\ref{fig5}C, we present a typical snapshot of the activator concentration profile along the cortex, with arrows indicating the magnitude and direction of the force generated by the Marangoni effect for the constrictive (purple) and dispersive (orange) cases. The force is largest at midpoints of the slopes on either side of the peak, where the gradient of $C_\activator$ is the largest. In the dispersive case, it acts to bulge out and widen the peak, whereas in the constrictive case to compress and sharpen it.

In the case of dispersive flow (negative $k_\sigma$), the Marangoni effect supports the diffusive transport of activator on the cortex. Indeed, the flow profiles in Fig~\ref{fig5}A and Fig~\ref{fig4} are similar and, if the coupling $k_\sigma$ is negative enough, we observe a similar behaviour of cells with a stable patch of activator in front and a slow straight movement (though with a slightly different shape of cortex cause by a different flow patterns).

In contrast, constrictive flow (positive $k_\sigma$) counteracts diffusion, effectively squeezing the patch, while the cytosolic flow carries bulk species away from the front and delivers them toward the sides of the cell. This makes the pattern of activator on cortex unstable and promotes oscillatory dynamics. To explain this effect it is useful to consider the oscillation as two wave packets travelling in opposite directions along the cortex. When these packets meet, they form a large peak; when separated, a trough forms between them. As shown in Fig~\ref{fig5}C, the constrictive force pushes the packets toward each other and, thus, increases their propagation speed. Consequently, constrictive flow accelerates the oscillation dynamics, reducing the oscillation period as the coupling strength increases (Fig~\ref{fig5}D). In contrast, dispersive flow slows oscillations at small negative $k_\sigma$ and upon further increase of the negative coupling, eventually, abolishes them entirely once a certain threshold of $k_\sigma$ is reached.

Rotation dynamics respond differently to the two flow types. Under constrictive flow, rotation transitions to oscillation once a positive coupling threshold is exceeded (around $0.03$ in Fig~\ref{fig5}D). This arises from the inherent asymmetry of the rotating peak: its leading edge (facing the direction of motion) is steeper than its trailing edge. This asymmetry generates a left--right force imbalance, with the constrictive flow preferentially accelerating waves travelling opposite to the motion of the peak. This rebalances the strengths of clockwise and counter-clockwise modes, ultimately shifting the pattern from rotation to oscillation. In contrast, strong dispersive flow affects rotation in the same way as oscillation, driving the system into static polarization once a negative coupling threshold is surpassed.

\subsection*{\label{sec:Class}Motility phenotypes}

\begin{figure}[h!]
\missingfigure{\includegraphics[width=\textwidth]{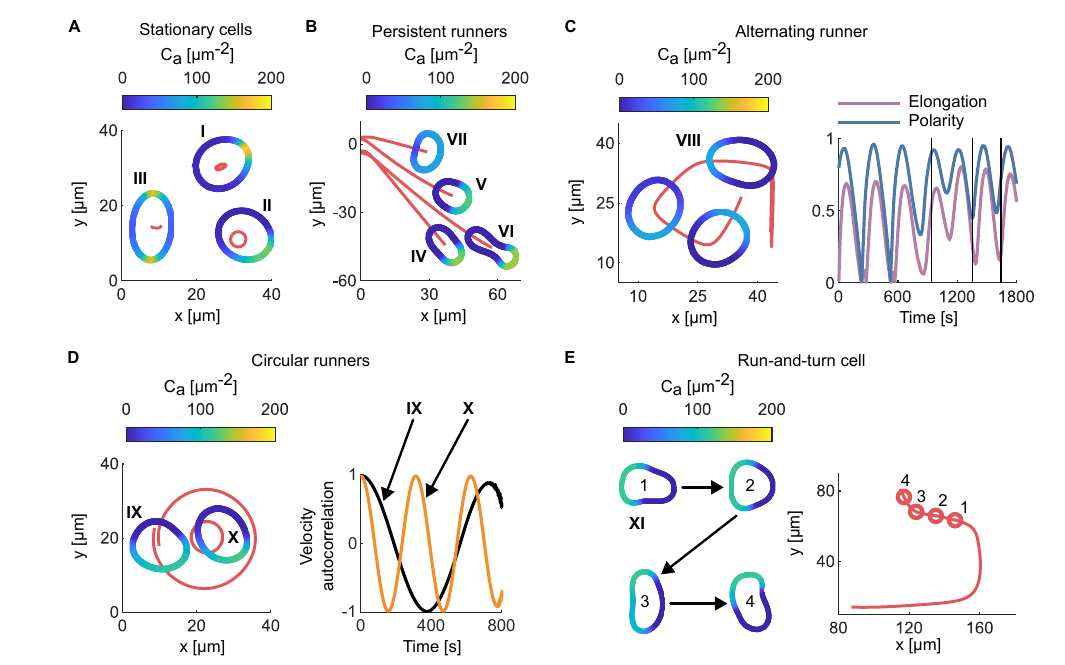}}%
\caption{\label{Fig:Class} \textbf{Characteristic motility phenotypes.} \textbf{A:} stationary cells, \textbf{B:} persistent runner cells, \textbf{C:} alternating runner cells, \textbf{D:} circular runner cells, and \textbf{E:} run-and-turn cells. In each panel snapshots of cells are presented with the colour code denoting the concentration of activator in the cortex (as shown by scale bars), and a trajectory of the centre of mass of the cell is plotted with red colour. In panel \textbf{C} we present the dependence of elongation and polarity on simulation time for the illustrated cell. Vertical black lines denote moments in which the plotted snapshots were taken. In panel \textbf{D} we plot velocity autocorrelation functions of presented cells. In panel \textbf{E} the trajectory is plotted separately on the right side and four circles mark the points in which snapshots plotted on the left side were taken. For the videos showing the time evolution of presented cells, see the \nameref{supporting_information} Section.
}
\end{figure}

In our simulations, by varying the strength of the couplings, number of proteins, and initial activator profiles, we have managed to generate a wide range of distinct behaviours of the cell. In order to describe them, it is useful to introduce a categorization into separate motility classes. In our analysis, we have used five distinct classes as presented in Fig~\ref{Fig:Class}.

\subsubsection*{\label{sec:Stationary}Stationary cells}

We start from the stationary cells, \ie, those that are practically not moving. We consider a cell to be stationary when its trajectory (the curve drawn by the geometrical centre) remains entirely inside the cell for the whole duration of the simulation.
Of course, the lack of movement does not imply that the cell reaches a stationary state. Typically, the concentration of activator on the cortex experience rotational (patch is moving around the cortex) or oscillatory (patch is disappearing and appearing on the cortex) dynamics which is accompanied with changes of the shape of cell. 

Fig~\ref{Fig:Class}A presents three examples of stationary cells. Cell I displays single patch oscillatory activator dynamics with a single patch appearing and disappearing alternately on opposite sides of the cell. This makes the shape of cell to resemble an egg with its centre oscillating along a trajectory that is almost a segment. Cell II presents rotational dynamics with the patch of activator moving around the cortex, which leads to a circular trajectory and an oval shape of the cell rotating along with the patch.

In our simulations, the stationary cells typically stay in the type of RD dynamics of the activator in which they were initialized. However, as shown in Fig~\ref{fig4}D, increase of the coupling $k_\sigma$ reduces a period of observed dynamics and favours oscillations over rotations. Another interesting example of changing of RD dynamics is shown by cell III in Fig~\ref{Fig:Class}A, where, by increasing the number of proteins, the cell was forced into oscillation activator dynamics with two simultaneously active regions. The cell is stretched by the patches alternately in vertical and horizontal direction making its shape elliptical with only a minimal motion of its centre of mass.

\subsubsection*{\label{sec:PersistentRunner}Persistent runner cells}

The second class of motility of simulated cells is the persistent runner. These cells have a stable pattern of activator in the cortex and move in a fixed direction; thus, within the accuracy of our simulations, they reach a stationary state. As discussed above, persistent runners emerge when the protrusive coupling $k_\mathrm{C}$ is large enough or when the surface tension coupling $k_\mathrm{\sigma}$ is negative enough, \ie, when there appear two vortices inside cell transporting species as shown in Figs.~\ref{fig4} and~\ref{fig5}B.

The interplay of these two distinct mechanisms leading to a persistent runner cells, manifests itself in a plethora of attained shapes. As shown in Fig~\ref{Fig:Class}B, depending on the values of coupling constants, the cell can have: a stadium geometry (cell IV); a fan shape with broader front and contracted rear (cell V); a dumbbell shape, in which the protrusive force is trying to extend the front while surface tension is resisting the elongation (cell VI); or a keratocyte-like shape with a
broad curved front and straight rear (cell VII).

\subsubsection*{\label{sec:AlternatingRunner}Alternating runner cells}

The third class, alternating runner cells, is characterised by oscillatory dynamics of activator proteins on the cortex giving rise to the characteristic motion with irregular turns separated by episodes of straight motion, as exemplified in Fig~\ref{Fig:Class}C (cell~VIII).  In this case, the patch of activator does not relocate itself to the exact opposite side of the cell, but fluctuates slightly with each oscillation producing a staggered motion, as can be seen from the trajectory in Fig~\ref{Fig:Class}C.

This process is also reflected in the magnitude of polarity as shown on the graph in Fig~\ref{Fig:Class}C: polarity drops during turn episodes when the entire cortex becomes active, and increases during run episodes. Furthermore, each run episode is accompanied by an increase in elongation of the cell body along the direction of motion. This elongation follows the peak of polarity expressed during a given run. Once the activator patch begins to relocate itself, the forces along the cortex become more evenly distributed, causing the cell to contract back toward more circular shape.

This class emerges naturally for simulations initiated with an oscillatory concentration profile with coupling parameters tuned to the crossover between stationary and persistent runners regions. However, it can also arise when simulations are initiated with a rotational profile with coupling constants at the edge of stability of rotation. In such cases, the cells typically undergo a transient phase, during which other characteristic behaviours may appear, before finally settling into alternating run.

\subsubsection*{\label{sec:CircularRunner}Circular runner cells}

The next class is the circular runners which groups cells that move along almost perfect circle rotating around a fixed point. This motion is always accompanied with the rotation dynamics of the patch of activator in the cortex. In our simulations we distinguish two variants of this class as presented in Fig~\ref{Fig:Class}D. The circular runner can have either a fan shape with the elongation direction almost perpendicular to the direction of motion (cell IX in Fig~\ref{Fig:Class}D) or a stadium shape with the elongation roughly parallel to the velocity (cell X in Fig~\ref{Fig:Class}D). Within this class, the fan-shaped cells are observed only for $k_\sigma<0$ while stadium-shaped appear exclusively for $k_\sigma>0$. Moreover, the radius of circular trajectory is much bigger for fun-shaped cells which makes their period of motion larger than for stadium-shaped cells, as shown in the plot of the velocity autocorrelation function in Fig~\ref{Fig:Class}D.

We note that, the circular runners of a stadium shape (cell X in Fig~\ref{Fig:Class}D) are quite similar to some of the stationary cells, like cell III in Fig~\ref{Fig:Class}A. Indeed, the main difference between these two groups is the radius of the circular trajectory which may or may not fit inside the cell. The distinguishing between them is an artifact of our (quite arbitrary) definition of stationary cells.

\subsubsection*{\label{sec:RunAndTurn}Run-and-turn cells}

The final class observed in our simulations are run-and-turn cells. It is observed for large values of $k_\mathrm{C}$ and negative $k_\sigma$, in which case the superposition of protrusive and dispersive flows makes the patch of activator on the cortex of the cell unstable. The resulting dynamics consist of periods of straight motion separated by turns, which justifies the name of this class.

Typical example of run-and-turn cell is presented in Fig~\ref{Fig:Class}E (cell XI). The cell appeared as a persistent runner for the first \SI{600}{\second} of simulation time, after which the turning began. As the cell moves, the initial fan shape of cortex (point 1 on the trajectory and snapshot 1 in Fig~\ref{Fig:Class}E) becomes more and more elongated. As a result, the cell slows down, and the patch of activator spreads (point 2 in Fig~\ref{Fig:Class}E) and, eventually, splits into two. Then, in a short time one of the new peaks starts to dominate (point 3 in Fig~\ref{Fig:Class}E). Lastly, the smaller peak completely disappears (point 4 in Fig~\ref{Fig:Class}E), making the cell turn and reform fan shape. The whole process then repeats itself---the cell XI has managed to make three turns over the simulation time.

We note that, run-and-turn cells and alternating runners have very similar trajectories. Nevertheless, the underlaying dynamics of activator is completely different---oscillatory for alternating runners and static with occasional lost of stability for run-and-turn class.

\section*{\label{sec:PD}Phase diagrams for varying coupling strengths}

After introducing the possible motility phenotypes and discussing the physical mechanisms behind them, we examine the characteristic behaviours that arise for different values of the coupling parameters $k_\mathrm{C}$ and $k_\mathrm{\sigma}$.

\begin{figure}[h!]
\missingfigure{\includegraphics[width=0.98\textwidth]{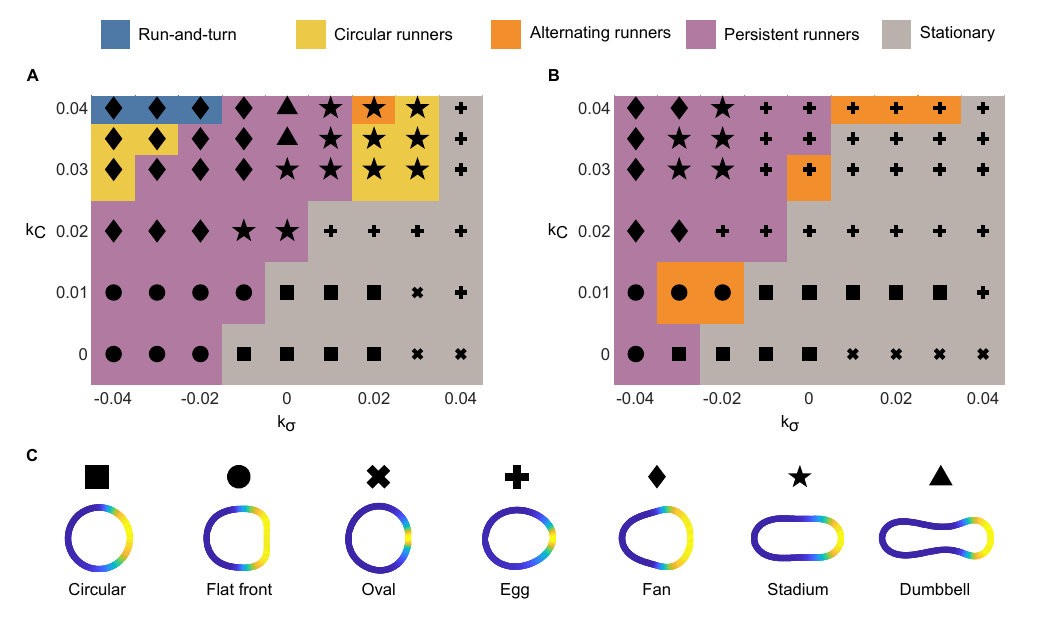}}%
\caption{\label{Fig:PD} \textbf{Phase diagrams.} Phase diagrams for limit cycle motility phenotypes as a function of coupling strength \textbf{A:} for cells initiated with a rotational concentration profile and \textbf{B:}  for cells initiated with an oscillatory concentration profile. Colour represents the motility class as shown above the panels, while marker represents characteristic shape of the cell. \textbf{C:} Legend specifying the characteristic shapes observed in the simulations.}
\end{figure}

Figure~\ref{Fig:PD} presents the results of a large number of simulations with systematically varied values of the coupling parameters. Moreover, each simulation was separately done with rotation (Fig~\ref{Fig:PD}A) and oscillation (Fig~\ref{Fig:PD}B) initial profiles of the activator. Each time, we have classified the motility phenotype after about \SI{1000}{\second} of simulation time and identified the shape of the cell assigning it to one of seven classes, as presented in Fig~\ref{Fig:PD}C. The characteristic shapes do not necessarily correspond to the end-of-simulation ones; rather, for each parameter pair, we select the contour that best represents the typical morphology over the simulated interval. We note that for some points, the assigned motility class might not be fully correct, as manifesting the asymptotic cycle may take much longer than the time of our simulation. 

Both diagrams are dominated by stationary and persistent runner cells. Stationary cells emerge when the protrusive coupling $k_\mathrm{C}$ is small and the surface tension coupling $k_\sigma$ is sufficiently strong. Such cells can take circular, oval, or egg shape and show very limited motion. Upon increasing $k_\mathrm{C}$ or reducing $k_\sigma$, the motility changes to the persistent runners class. These cells move along a line and, depending on the couplings, can take almost any possible shape.

An interesting behaviour is observed on the crossover between these two classes. For cells initialized with rotational state, upon leaving the domain of stationary class, typically the internal flows are able to move the cell but are not strong enough for a persistent motion. As a result, cells become circular runners. Alternatively, oscillatory cells in this region of parameters constantly switch between being stationary and persistent runners, which makes them fall into category of alternating runners.

Finally, for the highest considered values of $k_\mathrm{C}$ and for large negative $k_\sigma$, we discovered a region where internal flows inside the cell become strong enough to perturb the straight motion of the cell. In this case, the cell either moves over closed trajectory, becoming a circular runner, or makes random turns falling into the run-and-turn category. We were able to record these phenomena only for cells with a rotational initial state.

The observed dependence of the behaviour of the system on the initial state of the simulation can be explained with two distinct effects: First, since the oscillatory and rotational initial states differ in total protein numbers, their peak amplitudes also differ; consequently, larger coupling strengths are required for cells initiated from oscillation to achieve the same dynamics as for those initiated from rotation. The second key difference lies in the concentration profiles: the oscillatory state is symmetric, providing no built-in symmetry breaking, whereas the rotational state is inherently asymmetric. This lack of initial asymmetry in the oscillatory case affects the ability of cell to express certain behaviours (\textit{e.g.}, those that require directional bias) unless the couplings are sufficiently strong to spontaneously break this symmetry (with the help of numerical noise).

Our findings confirm indications of current state-of-the-art models and experiments~\cite{Liu2021,Eroume2021,Htet2025,Bruckner2024} that cytosolic flow, cell shape, and RD dynamics are tightly coupled in motile cells, jointly enabling a wide range of complex behaviours. Recent work also links signalling dynamics to energetic partitioning at the cortex~\cite{Chen2024}. Our formulation encapsulates these interacting processes and is consistent with this energy-partitioning perspective.

\section*{\label{sec:Con}Conclusion}
In this paper, we proposed a cross-scale mean-field theoretical framework to capture cellular locomotion for cells crawling on surfaces. By uniting intracellular signalling dynamics with spatial cytosolic flow and explicitly accounting for cortical and membrane surface flows, our model reproduces the wide range of experimentally observed cellular morphologies and motility patterns through small variations in coupling constants. Since these behaviours emerge as limit cycles, they reveal that coupling among these subsystems gives rise to intrinsically self-organised dynamics. Phase diagrams further demonstrate that the two types of coupling of the reaction–diffusion dynamics, with cytosolic flows---via protrusive force and spatially varying surface tension---are essential, along with diffusion dynamics, to recover the full repertoire of motility phenotypes observed in living cells. By systematically mapping the characteristic states of this coupled system, our work establishes a theoretical foundation that can inform and constrain future modelling and experimental efforts.

The current formulation is particularly suitable for investigating motility of cells lacking prominent contractile structures such as stress fibres, and amoeboid locomotion predominantly driven by actin polymerization~\cite{Lammermann2009}. However, faithfully capturing myosin-based contractility would likely require the inclusion of active stress coupling via a vector field representation, as discussed elsewhere~\cite{Moure2016}. Another natural extension involves the introduction of spatially variable viscosity between the external medium, the cortex, and the cytosol~\cite{Abreu2014}, which would enhance physical realism but demands a higher numerical resolution. Such a refinement would also allow the study of cortical ruffling phenomena arising from locally negative surface tension~\cite{Kantsler2007}. The current formulation and these extensions should ultimately be validated through a direct quantitative comparison with experiments, representing the next step for this framework.

In conclusion, our model highlights the critical role of two-dimensional cortical transport and the coupling between reaction–diffusion dynamics and membrane tension. This interplay generates Marangoni flows that, together with Fickian diffusion, mediate cross-talk between the cytosolic and cortical layers. As a result, microscopic membrane patterning becomes dynamically linked to across-cell signalling and hydrodynamic feedback, jointly governing cell shape, polarity, and locomotion patterns at the mesoscopic scale.

\section*{\label{supporting_information}Supporting information}

% Include only the SI item label in the paragraph heading. Use the \nameref{label} command to cite SI items in the text.
\paragraph*{S1 Appendix.}
\label{S1_Appendix}
\textbf{Details of the model and calculations.} Description of the details of the model, its numerical implementation, the method of analysis of the results, and parameters used in the simulations. 

%\paragraph*{S1 Video.}
%\label{S1_Video}
%\textbf{Cell I.}  Video presenting time evolution of cell I presented in Fig~\ref{Fig:Class}A.

%\paragraph*{S2 Video.}
%\label{S2_Video}
%\textbf{Cell II.}  Video presenting time evolution of cell II presented in Fig~\ref{Fig:Class}A.

%\paragraph*{S3 Video.}
%\label{S3_Video}
%\textbf{Cell III.}  Video presenting time evolution of cell III presented in Fig~\ref{Fig:Class}A.

%\paragraph*{S4 Video.}
%\label{S4_Video}
%\textbf{Cell IV.}  Video presenting time evolution of cell IV presented in Fig~\ref{Fig:Class}B.

%\paragraph*{S5 Video.}
%\label{S5_Video}
%\textbf{Cell V.}  Video presenting time evolution of cell V presented in Fig~\ref{Fig:Class}B.

%\paragraph*{S6 Video.}
%\label{S6_Video}
%\textbf{Cell VI.}  Video presenting time evolution of cell VI presented in Fig~\ref{Fig:Class}B.

%\paragraph*{S7 Video.}
%\label{S7_Video}
%\textbf{Cell VII.}  Video presenting time evolution of cell VII presented in Fig~\ref{Fig:Class}B.

%\paragraph*{S8 Video.}
%\label{S8_Video}
%\textbf{Cell VIII.}  Video presenting time evolution of cell VIII presented in Fig~\ref{Fig:Class}C.

%\paragraph*{S9 Video.}
%\label{S9_Video}
%\textbf{Cell IX.}  Video presenting time evolution of cell IX presented in Fig~\ref{Fig:Class}D.

%\paragraph*{S10 Video.}
%\label{S10_Video}
%\textbf{Cell X.}  Video presenting time evolution of cell X presented in Fig~\ref{Fig:Class}D.

%\paragraph*{S11 Video.}
%\label{S11_Video}
%\textbf{Cell XI.}  Video presenting time evolution of cell XI presented in Fig~\ref{Fig:Class}E.

\vspace{2mm}

The data presented in the figures, as well as the videos showing simulations in Fig~\ref{Fig:Class}, are available in the Zenodo archive \href{https://doi.org/10.5281/zenodo.17466623}{doi:10.5281/zenodo.17466623}.\\
The programme used for simulations is available on GitHub \url{https://github.com/blejzara/CellRDA-LS}.

\section*{Acknowledgments}

%The authors thank P. Jakubczyk, M. Napiórkowski, and A. Parry for inspiring discussions and suggestions. The paper has been supported by the German Research Foundation and the French National Research Agency project SM 289/8-1, AOBJ: 652939/ANR-18-CE92-0033-01, and by German Research Foundation project SM 289/10-1. A.M. was partially supported by the Polish National Science Center (Opus Grant No. 2022/45/B/ST3/00936).

BI thanks Marko Šoštar, Kevin Höllring, David Karatović, and Nikola Poljak for their insightful discussions and constructive feedback. PN is grateful to Anna Maciołek for valuable help.
IW acknowledges the support of the Croatian Science Foundation under project HRZZ-IP-2024-05-6331.
ASS has been supported by the German Research Foundation (DFG) under grant SM 289/10-1, as well as by the joint DFG---French National Research Agency (ANR) grant SM 289/11-1.

%\textcolor{green}{We thank Marko Šoštar, Kevin Höllring, David Karatović, and Nikola Poljak for their insightful discussions and constructive feedback, which substantially improved this work. We are also grateful to Tomislav Vuletić for general support throughout the project.}

\nolinenumbers

% Please compile your BiBTeX database using the "plos2025.bst" BibTeX style.
% This file is part of the current package.
% A sample BibTeX file is also included as "plos_bibtex_sample.bib".
%
% or
%
% Type in your references following Vancouver style and reference formatting instructions
% available at https://journals.plos.org/plosone/s/submission-guidelines#loc-references
% \begin{thebibliography}{}
% \bibitem{}
% Text
% \end{thebibliography}


\begin{thebibliography}{10}

\bibitem{Sackmann2010}
Sackmann E, Keber F, Heinrich D.
\newblock Physics of Cellular Movements.
\newblock Annual Review of Condensed Matter Physics. 2010 Aug;1:257-76.
\newblock \href {http://dx.doi.org/10.1146/annurev-conmatphys-070909-104105} {doi:10.1146/annurev-conmatphys-070909-104105}.

\bibitem{MerinoCasallo2022}
Merino-Casallo F, Gomez-Benito MJ, Hervas-Raluy S, Garcia-Aznar JM.
\newblock Unravelling cell migration: defining movement from the cell surface.
\newblock Cell Adhesion \& Migration. 2022 Dec;16(1):25-64.
\newblock \href {http://dx.doi.org/10.1080/19336918.2022.2055520} {doi:10.1080/19336918.2022.2055520}.

\bibitem{Mosaddeghzadeh2021}
Mosaddeghzadeh N, Ahmadian MR.
\newblock The RHO Family GTPases: Mechanisms of Regulation and Signaling.
\newblock Cells. 2021 Jul;10(7):1831.
\newblock \href {http://dx.doi.org/10.3390/cells10071831} {doi:10.3390/cells10071831}.

\bibitem{Jaffe2005}
Jaffe AB, Hall A.
\newblock Rho GTPases: biochemistry and biology.
\newblock Annual Review of Cell and Developmental Biology. 2005;21:247-69.
\newblock \href {http://dx.doi.org/10.1146/annurev.cellbio.21.020604.150721} {doi:10.1146/annurev.cellbio.21.020604.150721}.

\bibitem{Halatek2018}
Halatek J, Brauns F, Frey E.
\newblock Self-organization principles of intracellular pattern formation.
\newblock Philosophical Transactions of the Royal Society B: Biological Sciences. 2018;373(1747):20170107.
\newblock \href {http://dx.doi.org/10.1098/rstb.2017.0107} {doi:10.1098/rstb.2017.0107}.

\bibitem{Sostar2024}
\v{S}o\v{s}tar M, Marinovi\'{c} M, Fili\'{c} V, Pavin N, Weber I.
\newblock Oscillatory dynamics of Rac1 activity in \textit{Dictyostelium discoideum} amoebae.
\newblock PLOS Computational Biology. 2024;20(12):e1012025.
\newblock \href {http://dx.doi.org/10.1371/journal.pcbi.1012025} {doi:10.1371/journal.pcbi.1012025}.

\bibitem{Bement2024}
Bement WM, Goryachev AB, Miller AL, von Dassow G.
\newblock Patterning of the cell cortex by {Rho} {GTPases}.
\newblock Nature Reviews Molecular Cell Biology. 2024;25(4):290-308.
\newblock \href {http://dx.doi.org/10.1038/s41580-023-00682-z} {doi:10.1038/s41580-023-00682-z}.

\bibitem{Holmes2012}
Holmes WR, Edelstein-Keshet L.
\newblock A Comparison of Computational Models for Eukaryotic Cell Shape and Motility.
\newblock PLOS Computational Biology. 2012 December;8(12):e1002793.
\newblock \href {http://dx.doi.org/10.1371/journal.pcbi.1002793} {doi:10.1371/journal.pcbi.1002793}.

\bibitem{Ziebert2016}
Ziebert F, Aranson IS.
\newblock Computational approaches to substrate-based cell motility.
\newblock npj Computational Materials. 2016 July;2:16019.
\newblock \href {http://dx.doi.org/10.1038/npjcompumats.2016.19} {doi:10.1038/npjcompumats.2016.19}.

\bibitem{Buttenschon2020}
Buttensch{\"o}n A, Edelstein-Keshet L.
\newblock Bridging from single to collective cell migration: A review of models and links to experiments.
\newblock PLOS Computational Biology. 2020 December;16(12):e1008411.
\newblock \href {http://dx.doi.org/10.1371/journal.pcbi.1008411} {doi:10.1371/journal.pcbi.1008411}.

\bibitem{Turing1952}
Turing AM.
\newblock The Chemical Basis of Morphogenesis.
\newblock Philosophical Transactions of the Royal Society of London Series B, Biological Sciences. 1952 August;237(641):37-72.
\newblock \href {http://dx.doi.org/10.1098/rstb.1952.0012} {doi:10.1098/rstb.1952.0012}.

\bibitem{Beta2017}
Beta C, Kruse K.
\newblock Intracellular Oscillations and Waves.
\newblock Annual Review of Condensed Matter Physics. 2017;8:239-64.
\newblock \href {http://dx.doi.org/10.1146/annurev-conmatphys-031016-025210} {doi:10.1146/annurev-conmatphys-031016-025210}.

\bibitem{Jilkine2007}
Jilkine A, Marée AFM, Edelstein-Keshet L.
\newblock Mathematical Model for Spatial Segregation of the Rho-Family GTPases Based on Inhibitory Crosstalk.
\newblock Bulletin of Mathematical Biology. 2007 August;69(6):1943-78.
\newblock \href {http://dx.doi.org/10.1007/s11538-007-9200-6} {doi:10.1007/s11538-007-9200-6}.

\bibitem{Otsuji2007}
Otsuji M, Ishihara S, Co C, Kaibuchi K, Mochizuki A, Kuroda S.
\newblock A Mass Conserved Reaction–Diffusion System Captures Properties of Cell Polarity.
\newblock PLOS Computational Biology. 2007 June;3(6):e108.
\newblock \href {http://dx.doi.org/10.1371/journal.pcbi.0030108} {doi:10.1371/journal.pcbi.0030108}.

\bibitem{Bement2015}
Bement WM, Leda M, Moe AM, Kita AM, Larson ME, Golding AE, et~al.
\newblock Activator–inhibitor coupling between Rho signalling and actin assembly makes the cell cortex an excitable medium.
\newblock Nature Cell Biology. 2015 November;17(11):1471-83.
\newblock \href {http://dx.doi.org/10.1038/ncb3251} {doi:10.1038/ncb3251}.

\bibitem{Chiou2018}
Chiou JG, Ramirez SA, Elston TC, Witelski TP, Schaeffer DG, Lew DJ.
\newblock Principles that govern competition or co-existence in Rho-GTPase driven polarization.
\newblock PLOS Computational Biology. 2018;14(4):e1006095.
\newblock \href {http://dx.doi.org/10.1371/journal.pcbi.1006095} {doi:10.1371/journal.pcbi.1006095}.

\bibitem{Neilson2010}
Neilson MP, Mackenzie JA, Webb SD, Insall RH.
\newblock Use of the parameterised finite element method to robustly and efficiently evolve the edge of a moving cell.
\newblock Integrative Biology. 2010 November;2(11-12):687-95.
\newblock \href {http://dx.doi.org/10.1039/c0ib00047g} {doi:10.1039/c0ib00047g}.

\bibitem{MacDonald2016}
MacDonald G, Mackenzie JA, Nolan M, Insall RH.
\newblock A computational method for the coupled solution of reaction--diffusion equations on evolving domains and manifolds: Application to a model of cell migration and chemotaxis.
\newblock Journal of Computational Physics. 2016 April;309:207-26.
\newblock \href {http://dx.doi.org/10.1016/j.jcp.2015.12.038} {doi:10.1016/j.jcp.2015.12.038}.

\bibitem{Mackenzie2019}
Mackenzie JA, Rowlatt CF, Insall RH.
\newblock A Conservative Finite Element ALE Scheme for Mass-Conserving Reaction-Diffusion Equations on Evolving Two-Dimensional Domains.
\newblock arXiv preprint arXiv:191002282. 2019 October.
\newblock \href {http://dx.doi.org/10.48550/arXiv.1910.02282} {doi:10.48550/arXiv.1910.02282}.

\bibitem{Maree2012}
Marée AFM, Grieneisen VA, Edelstein-Keshet L.
\newblock How Cells Integrate Complex Stimuli: The Effect of Feedback from Phosphoinositides and Cell Shape on Cell Polarization and Motility.
\newblock PLOS Computational Biology. 2012 March;8(3):e1002402.
\newblock \href {http://dx.doi.org/10.1371/journal.pcbi.1002402} {doi:10.1371/journal.pcbi.1002402}.

\bibitem{Aranson2016}
Aranson IS, editor.
\newblock Physical Models of Cell Motility.
\newblock Biological and Medical Physics, Biomedical Engineering. Cham, Switzerland: Springer; 2016.
\newblock \href {http://dx.doi.org/10.1007/978-3-319-24448-8} {doi:10.1007/978-3-319-24448-8}.

\bibitem{Shao2012}
Shao D, Levine H, Rappel WJ.
\newblock Coupling actin flow, adhesion, and morphology in a computational cell motility model.
\newblock Proceedings of the National Academy of Sciences. 2012 May;109(18):6851-6.
\newblock \href {http://dx.doi.org/10.1073/pnas.1203252109} {doi:10.1073/pnas.1203252109}.

\bibitem{Alonso2018}
Alonso S, Stange M, Beta C.
\newblock Modeling random crawling, membrane deformation and intracellular polarity of motile amoeboid cells.
\newblock PLOS ONE. 2018 August;13(8):e0201977.
\newblock \href {http://dx.doi.org/10.1371/journal.pone.0201977} {doi:10.1371/journal.pone.0201977}.

\bibitem{Moreno2020}
Moreno E, Flemming S, Font F.
\newblock Modeling cell crawling strategies with a bistable model: From amoeboid to fan-shaped cell motion.
\newblock Physica D: Nonlinear Phenomena. 2020 November;412:132591.
\newblock \href {http://dx.doi.org/10.1016/j.physd.2020.132591} {doi:10.1016/j.physd.2020.132591}.

\bibitem{Moreno2022a}
Moreno E, Großmann R, Beta C, Alonso S.
\newblock From Single to Collective Motion of Social Amoebae: A Computational Study of Interacting Cells.
\newblock Frontiers in Physics. 2022 February;9:750187.
\newblock \href {http://dx.doi.org/10.3389/fphy.2021.750187} {doi:10.3389/fphy.2021.750187}.

\bibitem{Osher1988}
Osher S, Sethian JA.
\newblock Fronts Propagating with Curvature-Dependent Speed: Algorithms Based on Hamilton--Jacobi Formulations.
\newblock Journal of Computational Physics. 1988;79(1):12-49.
\newblock \href {http://dx.doi.org/10.1016/0021-9991(88)90002-2} {doi:10.1016/0021-9991(88)90002-2}.

\bibitem{Kuusela2009}
Kuusela E, Alt W.
\newblock Continuum model of cell adhesion and migration.
\newblock Journal of Mathematical Biology. 2009 January;58(1-2):135-61.
\newblock \href {http://dx.doi.org/10.1007/s00285-008-0179-x} {doi:10.1007/s00285-008-0179-x}.

\bibitem{Shi2013}
Shi C, Huang CH, Devreotes PN, Iglesias PA.
\newblock Interaction of Motility, Directional Sensing, and Polarity Modules Recreates the Behaviors of Chemotaxing Cells.
\newblock PLOS Computational Biology. 2013 July;9(7):e1003122.
\newblock \href {http://dx.doi.org/10.1371/journal.pcbi.1003122} {doi:10.1371/journal.pcbi.1003122}.

\bibitem{Schindler2024}
Schindler D, Moldenhawer T, Beta C, Huisinga W, Holschneider M.
\newblock Three-component contour dynamics model to simulate and analyze amoeboid cell motility in two dimensions.
\newblock PLOS ONE. 2024 January;19(1):e0297511.
\newblock \href {http://dx.doi.org/10.1371/journal.pone.0297511} {doi:10.1371/journal.pone.0297511}.

\bibitem{Moure2016}
Moure A, Gomez H.
\newblock Computational model for amoeboid motion: Coupling membrane and cytosol dynamics.
\newblock Physical Review E. 2016 October;94(4):042423.
\newblock \href {http://dx.doi.org/10.1103/PhysRevE.94.042423} {doi:10.1103/PhysRevE.94.042423}.

\bibitem{Bruckner2024}
Br{\"u}ckner DB, Broedersz CP.
\newblock Learning dynamical models of single and collective cell migration: a review.
\newblock Reports on Progress in Physics. 2024;87(5):056601.
\newblock \href {http://dx.doi.org/10.1088/1361-6633/ad36d2} {doi:10.1088/1361-6633/ad36d2}.

\bibitem{Osher2003}
Osher S, Fedkiw RP.
\newblock Level Set Methods and Dynamic Implicit Surfaces.
\newblock 1st ed. No. 153 in Applied Mathematical Sciences. New York, NY: Springer-Verlag; 2003.
\newblock Reviewed in *Applied Mechanics Reviews* 57 (3): B15 (2004), doi:10.1115/1.1760520.
\newblock \href {http://dx.doi.org/10.1007/b98879} {doi:10.1007/b98879}.

\bibitem{Anderson1998}
Anderson DM, McFadden GB, Wheeler AA.
\newblock Diffuse-Interface Methods in Fluid Mechanics.
\newblock Annual Review of Fluid Mechanics. 1998;30:139-65.
\newblock \href {http://dx.doi.org/10.1146/annurev.fluid.30.1.139} {doi:10.1146/annurev.fluid.30.1.139}.

\bibitem{Peskin2002}
Peskin CS.
\newblock The Immersed Boundary Method.
\newblock Acta Numerica. 2002;11:479-517.
\newblock \href {http://dx.doi.org/10.1017/S0962492902000077} {doi:10.1017/S0962492902000077}.

\bibitem{Russo2000}
Russo G, Smereka P.
\newblock A Remark on Computing Distance Functions.
\newblock Journal of Computational Physics. 2000;163(1):51-67.
\newblock \href {http://dx.doi.org/10.1006/jcph.2000.6553} {doi:10.1006/jcph.2000.6553}.

\bibitem{Parameswaran2023}
Parameswaran S, Mandal JC.
\newblock A stable interface-preserving reinitialization equation for conservative level set method.
\newblock European Journal of Mechanics – B/Fluids. 2023;98:40-63.
\newblock \href {http://dx.doi.org/10.1016/j.euromechflu.2022.11.001} {doi:10.1016/j.euromechflu.2022.11.001}.

\bibitem{Landau1987}
Landau LD, Lifshitz EM.
\newblock Fluid Mechanics.
\newblock 2nd ed. No.~6 in Course of Theoretical Physics. Oxford: Pergamon Press; 1987.
\newblock \href {http://dx.doi.org/10.1016/C2013-0-03799-1} {doi:10.1016/C2013-0-03799-1}.

\bibitem{Brackbill1992}
Brackbill JU, Kothe DB, Zemach C.
\newblock A Continuum Method for Modeling Surface Tension.
\newblock Journal of Computational Physics. 1992;100(2):335-54.
\newblock \href {http://dx.doi.org/10.1016/0021-9991(92)90240-Y} {doi:10.1016/0021-9991(92)90240-Y}.

\bibitem{Whitfield2016}
Whitfield CA, Hawkins RJ.
\newblock Instabilities, motion and deformation of active fluid droplets.
\newblock New Journal of Physics. 2016 December;18(12):123016.
\newblock \href {http://dx.doi.org/10.1088/1367-2630/18/12/123016} {doi:10.1088/1367-2630/18/12/123016}.

\bibitem{Mogilner2003}
Mogilner A, Oster G.
\newblock Force Generation by Actin Polymerization {II}: The Elastic Ratchet and Tethered Filaments.
\newblock Biophysical Journal. 2003;84(3):1591-605.
\newblock \href {http://dx.doi.org/10.1016/S0006-3495(03)74969-8} {doi:10.1016/S0006-3495(03)74969-8}.

\bibitem{Barnhart2011}
Barnhart EL, Lee KC, Keren K, Mogilner A, Theriot JA.
\newblock An Adhesion‐Dependent Switch between Mechanisms That Determine Motile Cell Shape.
\newblock PLOS Biology. 2011;9(5):e1001059.
\newblock \href {http://dx.doi.org/10.1371/journal.pbio.1001059} {doi:10.1371/journal.pbio.1001059}.

\bibitem{Fogelson2015}
Fogelson AL, Keener JP.
\newblock A Framework for Exploring the Post-gelation Behavior of Ziff and Stell's Polymerization Models.
\newblock SIAM Journal on Applied Mathematics. 2015;75(3):1346-68.
\newblock \href {http://dx.doi.org/10.1137/140983872} {doi:10.1137/140983872}.

\bibitem{Chen2015}
Chen J, Yang C, Mao ZS.
\newblock The interphase mass transfer in liquid--liquid systems with Marangoni effect.
\newblock The European Physical Journal Special Topics. 2015 Mar;224(2):389-99.
\newblock \href {http://dx.doi.org/10.1140/epjst/e2015-02368-0} {doi:10.1140/epjst/e2015-02368-0}.

\bibitem{Schmitt2016}
Schmitt M, Stark H.
\newblock Marangoni flow at droplet interfaces: Three-dimensional solution and applications.
\newblock Physics of Fluids. 2016 01;28(1):012106.
\newblock \href {http://dx.doi.org/10.1063/1.4939212} {doi:10.1063/1.4939212}.

\bibitem{Merkel2000}
Merkel R, Simson R, Simson DA, Hohenadl M, Boulbitch A, Wallraff E, et~al.
\newblock A Micromechanic Study of Cell Polarity and Plasma Membrane--Cell Body Coupling in \emph{Dictyostelium}.
\newblock Biophysical Journal. 2000 Aug;79(2):707-19.
\newblock \href {http://dx.doi.org/10.1016/S0006-3495(00)76329-6} {doi:10.1016/S0006-3495(00)76329-6}.

\bibitem{Clark2014}
Clark AG, Wartlick O, Salbreux G, Paluch EK.
\newblock Stresses at the cell surface during animal cell morphogenesis.
\newblock Curr Biol. 2014;24(10):R484-94.
\newblock \href {http://dx.doi.org/10.1016/j.cub.2014.03.059} {doi:10.1016/j.cub.2014.03.059}.

\bibitem{Chugh2017}
Chugh P, Clark AG, Smith MB, Cassani DAD, Dierkes K, Ragab A, et~al.
\newblock Actin cortex architecture regulates cell surface tension.
\newblock Nature Cell Biology. 2017 Jun;19(6):689-97.
\newblock \href {http://dx.doi.org/10.1038/ncb3525} {doi:10.1038/ncb3525}.

\bibitem{Li2024}
Li M, Xing X, Yuan J, Zeng Z.
\newblock Research progress on the regulatory role of cell membrane surface tension in cell behavior.
\newblock Heliyon. 2024;10(9):e29923.
\newblock \href {http://dx.doi.org/10.1016/j.heliyon.2024.e29923} {doi:10.1016/j.heliyon.2024.e29923}.

\bibitem{GarciaArcos2024}
Garc{\'\i}a-Arcos JM, Mehidi A, S{\'a}nchez~Vel{\'a}zquez J, Guillamat P, Tomba C, Houzet L, et~al.
\newblock Actin dynamics sustains spatial gradients of membrane tension in adherent cells.
\newblock bioRxiv. 2024.
\newblock Preprint.
\newblock \href {http://dx.doi.org/10.1101/2024.07.15.603517} {doi:10.1101/2024.07.15.603517}.

\bibitem{Mogilner1996}
Mogilner A, Oster G.
\newblock Cell motility driven by actin polymerization.
\newblock Biophysical Journal. 1996 December;71(6):3030-45.
\newblock \href {http://dx.doi.org/10.1016/S0006-3495(96)79496-1} {doi:10.1016/S0006-3495(96)79496-1}.

\bibitem{Scriven1960}
Scriven LE, Sternling CV.
\newblock The Marangoni Effects.
\newblock Nature. 1960;187(4733):186-8.
\newblock \href {http://dx.doi.org/10.1038/187186a0} {doi:10.1038/187186a0}.

\bibitem{Sens2015}
Sens P, Plastino J.
\newblock Membrane tension and cytoskeleton organization in cell motility.
\newblock Journal of Physics: Condensed Matter. 2015;27(27):273103.
\newblock \href {http://dx.doi.org/10.1088/0953-8984/27/27/273103} {doi:10.1088/0953-8984/27/27/273103}.

\bibitem{Machacek2009}
Machacek M, Hodgson L, Welch C, Elliott H, Pertz O, Nalbant P, et~al.
\newblock Coordination of Rho GTPase activities during cell protrusion.
\newblock Nature. 2009 Sep;461(7260):99-103.
\newblock Epub 2009 Aug 19.
\newblock \href {http://dx.doi.org/10.1038/nature08242} {doi:10.1038/nature08242}.

\bibitem{Houk2012}
Houk AR, Jilkine A, Mejean CO, Boltyanskiy R, Dufresne ER, Angenent SB, et~al.
\newblock Membrane tension maintains cell polarity by confining signals to the leading edge during neutrophil migration.
\newblock Cell. 2012 Jan;148(1--2):175-88.
\newblock \href {http://dx.doi.org/10.1016/j.cell.2011.10.050} {doi:10.1016/j.cell.2011.10.050}.

\bibitem{Goryachev2019}
Goryachev AB, Leda M.
\newblock Autoactivation of small GTPases by the GEF–effector positive feedback modules.
\newblock F1000Research. 2019;8:1676.
\newblock \href {http://dx.doi.org/10.12688/f1000research.20003.1} {doi:10.12688/f1000research.20003.1}.

\bibitem{Sostar2022}
\v{S}o\v{s}tar M.
\newblock Analiza dinamike proteina Rac1 tijekom staničnog kretanja [PhD thesis].
\newblock Zagreb, Hrvatska: Sveučilište u Zagrebu, Prirodoslovno-matematički fakultet; 2022.
\newblock Disertacija, otvoreni pristup.
\newblock Available from: \url{https://urn.nsk.hr/urn:nbn:hr:217:457625}.

\bibitem{Das2012}
Das M, Drake T, Wiley DJ, Buchwald P, Vavylonis D, Verde F.
\newblock Oscillatory dynamics of {Cdc42} {GTPase} in the control of polarized growth.
\newblock Science. 2012;337(6091):239-43.
\newblock \href {http://dx.doi.org/10.1126/science.1218377} {doi:10.1126/science.1218377}.

\bibitem{Graessl2017}
Graessl M, Koch J, Calderon A, Kamps D, Banerjee S, Mazel T, et~al.
\newblock An excitable Rho GTPase signaling network generates dynamic subcellular contraction patterns.
\newblock Journal of Cell Biology. 2017;216(12):4271-85.
\newblock \href {http://dx.doi.org/10.1083/jcb.201706052} {doi:10.1083/jcb.201706052}.

\bibitem{Michaux2018}
Michaux JB, Robin FB, McFadden WM, Munro EM.
\newblock Excitable RhoA dynamics drive pulsed contractions in the early \textit{C.~elegans} embryo.
\newblock Journal of Cell Biology. 2018;217(12):4230-52.
\newblock \href {http://dx.doi.org/10.1083/jcb.201806161} {doi:10.1083/jcb.201806161}.

\bibitem{Mori2008}
Mori Y, Jilkine A, Edelstein-Keshet L.
\newblock Wave-Pinning and Cell Polarity from a Bistable Reaction–Diffusion System.
\newblock Biophysical Journal. 2008;94(9):3684-97.
\newblock \href {http://dx.doi.org/10.1529/biophysj.107.120824} {doi:10.1529/biophysj.107.120824}.

\bibitem{Douglass2005}
Douglass AD, Vale RD.
\newblock Single‐Molecule Microscopy Reveals Plasma Membrane Microdomains Created by Protein‐Protein Networks That Exclude or Trap Signaling Molecules in T Cells.
\newblock Cell. 2005;121(6):937-50.
\newblock \href {http://dx.doi.org/10.1016/j.cell.2005.04.009} {doi:10.1016/j.cell.2005.04.009}.

\bibitem{Teigen2011}
Teigen KE, Song P, Lowengrub J, Voigt A.
\newblock A diffuse-interface method for two-phase flows with soluble surfactants.
\newblock J Comput Phys. 2011;230(2):375-93.
\newblock \href {http://dx.doi.org/10.1016/j.jcp.2010.09.020} {doi:10.1016/j.jcp.2010.09.020}.

\bibitem{Harlow1965}
Harlow FH, Welch JE.
\newblock Numerical Calculation of Time-Dependent Viscous Incompressible Flow of Fluid with Free Surface.
\newblock Physics of Fluids. 1965;8(12):2182-9.
\newblock \href {http://dx.doi.org/10.1063/1.1761178} {doi:10.1063/1.1761178}.

\bibitem{Chorin1968}
Chorin A.
\newblock Numerical Solution of the Navier--Stokes Equations.
\newblock Mathematics of Computation. 1968;22(104):745-62.
\newblock \href {http://dx.doi.org/10.1090/S0025-5718-1968-0242392-2} {doi:10.1090/S0025-5718-1968-0242392-2}.

\bibitem{Oono1988}
Oono Y, Puri S.
\newblock Study of Phase-Separation Dynamics by Use of Cell Dynamical Systems. I. Modeling.
\newblock Physical Review A. 1988;38(1):434-53.
\newblock \href {http://dx.doi.org/10.1103/PhysRevA.38.434} {doi:10.1103/PhysRevA.38.434}.

\bibitem{C99}
{International Organization for Standardization (ISO)}, {International Electrotechnical Commission (IEC)}. Information technology --- Programming languages --- C; 2024.
\newblock Current C standard (C23). ISO does not provide a DOI.
\newblock International Standard {ISO/IEC 9899:2024}, 5th ed.
\newblock Available from: \url{https://www.iso.org/standard/82075.html}.

\bibitem{Fiji}
Schindelin J, Arganda-Carreras I, Frise E, Kaynig V, Longair M, Pietzsch T, et~al.
\newblock Fiji: An Open-Source Platform for Biological-Image Analysis.
\newblock Nature Methods. 2012;9(7):676-82.
\newblock Fiji/ImageJ v1.54p was used in this study.
\newblock \href {http://dx.doi.org/10.1038/nmeth.2019} {doi:10.1038/nmeth.2019}.

\bibitem{Quimp}
Baniukiewicz P, Collier S, Bretschneider T.
\newblock QuimP: Analyzing Transmembrane Signalling in Highly Deformable Cells.
\newblock Bioinformatics. 2018;34(15):2695-7.
\newblock QuimP module for Fiji/ImageJ (accessed 23 May 2025).
\newblock \href {http://dx.doi.org/10.1093/bioinformatics/bty169} {doi:10.1093/bioinformatics/bty169}.

\bibitem{MATLAB2023b}
{The MathWorks, Inc }. {MATLAB} (R2023b). Natick, Massachusetts; 2023.
\newblock [Computer software].
\newblock Available from: \url{https://www.mathworks.com/products/matlab.html}.

\bibitem{Dunn1990}
Dunn GA, Brown AF.
\newblock Quantifying Cellular Shape Using Moment Invariants.
\newblock In: Alt W, Hoffmann G, editors. Biological Motion. vol.~89 of Lecture Notes in Biomathematics. Berlin, Heidelberg: Springer; 1990. p. 10-34.
\newblock \href {http://dx.doi.org/10.1007/978-3-642-51664-1\_2} {doi:10.1007/978-3-642-51664-1\_2}.

\bibitem{Teague1980}
Teague MR.
\newblock Image Analysis via the General Theory of Moments.
\newblock Journal of the Optical Society of America. 1980 Aug;70(8):920-30.
\newblock \href {http://dx.doi.org/10.1364/JOSA.70.000920} {doi:10.1364/JOSA.70.000920}.

\bibitem{Niwayama2016}
Niwayama R, Nagao H, Kitajima TS, Hufnagel L, Shinohara K, Higuchi T, et~al.
\newblock Bayesian Inference of Forces Causing Cytoplasmic Streaming in \textit{Caenorhabditis elegans} Embryos and Mouse Oocytes.
\newblock PLOS ONE. 2016;11(7):e0159917.
\newblock \href {http://dx.doi.org/10.1371/journal.pone.0159917} {doi:10.1371/journal.pone.0159917}.

\bibitem{Htet2025}
Htet PH, Lauga E.
\newblock Analytical methods for cytoplasmic streaming in elongated cells.
\newblock PNAS Nexus. 2025;4(3):pgaf057.
\newblock \href {http://dx.doi.org/10.1093/pnasnexus/pgaf057} {doi:10.1093/pnasnexus/pgaf057}.

\bibitem{Illukkumbura2020}
Illukkumbura R, Bland T, Goehring NW.
\newblock Patterning and polarization of cells by intracellular flows.
\newblock Current Opinion in Cell Biology. 2020;62:123-34.
\newblock \href {http://dx.doi.org/10.1016/j.ceb.2019.10.005} {doi:10.1016/j.ceb.2019.10.005}.

\bibitem{Meindlhumer2023}
Meindlhumer S, Brauns F, Fin{\v z}gar JR, Kerssemakers J, Dekker C, Frey E.
\newblock Directing Min protein patterns with advective bulk flow.
\newblock Nat Comm. 2023;14(1):450.
\newblock \href {http://dx.doi.org/10.1038/s41467-023-35997-0} {doi:10.1038/s41467-023-35997-0}.

\bibitem{Wigbers2020}
Wigbers MC, Brauns F, Leung CY, Frey E.
\newblock Flow induced symmetry breaking in a conceptual polarity model.
\newblock Cells. 2020;9(6):1524.
\newblock \href {http://dx.doi.org/10.3390/cells9061524} {doi:10.3390/cells9061524}.

\bibitem{Chugh2018}
Chugh P, Paluch EK.
\newblock The actin cortex at a glance.
\newblock Journal of Cell Science. 2018;131(14):jcs186254.
\newblock Review.
\newblock \href {http://dx.doi.org/10.1242/jcs.186254} {doi:10.1242/jcs.186254}.

\bibitem{Liu2021}
Liu Y, Rens EG, Edelstein-Keshet L.
\newblock Spots, stripes, and spiral waves in models for static and motile cells.
\newblock Journal of Mathematical Biology. 2021;82(4):28.
\newblock \href {http://dx.doi.org/10.1007/s00285-021-01550-0} {doi:10.1007/s00285-021-01550-0}.

\bibitem{Eroume2021}
Eroum{\'e} KS, Vasilevich A, Vermeulen S, de~Boer J, Carlier A.
\newblock On the influence of cell shape on dynamic reaction--diffusion polarization patterns.
\newblock PLOS ONE. 2021;16(3):e0248293.
\newblock \href {http://dx.doi.org/10.1371/journal.pone.0248293} {doi:10.1371/journal.pone.0248293}.

\bibitem{Chen2024}
Chen S, Seara DS, Michaud A, Kim S, Bement WM, Murrell MP.
\newblock Energy partitioning in the cell cortex.
\newblock Nature Physics. 2024;20(11):1824-32.
\newblock \href {http://dx.doi.org/10.1038/s41567-024-02626-6} {doi:10.1038/s41567-024-02626-6}.

\bibitem{Lammermann2009}
L{\"a}mmermann T, Sixt M.
\newblock Mechanical modes of 'amoeboid' cell migration.
\newblock Curr Opin Cell Biol. 2009;21(5):636-44.
\newblock \href {http://dx.doi.org/10.1016/j.ceb.2009.05.003} {doi:10.1016/j.ceb.2009.05.003}.

\bibitem{Abreu2014}
Abreu D, Levant M, Steinberg V, Seifert U.
\newblock Fluid vesicles in flow.
\newblock Adv Colloid Interface Sci. 2014;208:129-41.
\newblock \href {http://dx.doi.org/10.1016/j.cis.2014.02.004} {doi:10.1016/j.cis.2014.02.004}.

\bibitem{Kantsler2007}
Kantsler V, Segre E, Steinberg V.
\newblock Vesicle Dynamics in Time-Dependent Elongation Flow: Wrinkling Instability.
\newblock Phys Rev Lett. 2007;99(17):178102.
\newblock \href {http://dx.doi.org/10.1103/PhysRevLett.99.178102} {doi:10.1103/PhysRevLett.99.178102}.

\end{thebibliography}
\end{document}